\documentclass{appolb}
\usepackage{graphicx}

\begin{document}
\title{Higher order moments of net-charge and multiplicity
distributions in p+p interactions at SPS energies from
NA61/SHINE
\thanks{Presented at the Critical Point and Onset of Deconfinement 2016, Wroclaw, Poland}%
}
\author{Maja Ma\'{c}kowiak-Paw{\l}owska for the NA61/SHINE Collaboration
\address{Faculty of Physics, Warsaw University of Technology, Koszykowa 75, 00-662 Warsaw, Poland}
}
\maketitle
\begin{abstract}
NA61/SHINE at the CERN SPS is a fixed-target experiment pursuing a rich physics program including measurements for heavy ion, neutrino and cosmic ray physics. The main goal of the ion program is to study the properties of the onset of deconfinement and to search for the signatures of the critical point. A specific property of the critical point, the increase in the correlation length, makes fluctuations its basic signal. Higher order moments of suitable observables are of special interest as they are more sensitive to the correlation length than typically studied second order moments.

In this contribution preliminary results on higher order fluctuations of negatively charged hadron multiplicity and net-charge in p+p interactions will be shown. The new data will be compared with model predictions.
\end{abstract}
\PACS{25.75.-q, 25.75.Gz, 25.75.Nq}
  
\section{Introduction}
One of the most important goals of high-energy heavy-ion collisions is to establish the phase diagram of strongly interacting matter by finding the possible phase boundaries and critical points. A specific property of the critical point, the increase in the correlation length $\xi$, makes 
fuctuations its basic signal~\cite{Stephanov_overview}. Fluctuations are quantified by moments of measured distributions of suitable observables of order higher than the first.

Critical point fluctuations are expected to increase as (approximately) $\xi^{2}$ for the  variance (second moment) of event-by-event observables such as multiplicities or mean transverse momenta of particles. Higher, non-Gaussian, moments of fluctuations should depend more sensitively
on $\xi$, e.g. the fourth moment is expected to grow as $\xi^{7}$ near the critical point~\cite{Stephanov:2011zz}.

This contribution shows results on fluctuations of negatively charged hadron and net-charge multiplicity distributions defined by moments or cumulants up to the fourth moment. Net-charge is defined as the difference between positively and negatively charged hadron multiplicities in p+p interactions collected by the NA61/SHINE experiment~\cite{Antoniou:2006mh} in 2009. The reason to focus on multiplicity of negatively charged particle is the fact that the underlying correlations are almost insensitive to
resonance decays as there are very few resonances decaying into pairs of negatively charged particles. Net-charge, under some assumptions allows to compare data to QCD calculations on lattice.
\section{Fluctuation measures}
In the grand canonical ensemble mean, variance and in general cumulants (denoted with index ${c}$) of a multiplicity distribution are extensive quantities~(they are proportionl to volume $\sim V$).
A ratio of two extensive quantities is an intensive quantity e.g.:
		\begin{center}
			$\omega[N] = \frac{Var[N]}{\langle N \rangle}$,
		\end{center}
		where $Var[N]$ and $\langle N \rangle$ are variance and mean of the multiplicity distribution. The scaled variance is independent of $V$ (for event ensembles with fixed $V$) but it depends on fluctuations of $V$ (even if $\langle V \rangle$ is fixed). For the Poisson distribution (independent particle production) $\omega=1$.
		
	For third and fourth order cumulants there are several possibilities for deriving intensive measures. The two most popular are:
\begin{equation}
	\frac{\langle N^{3}\rangle_{c}}{Var[N]},\quad
	\frac{\langle N^{4}\rangle_{c}}{Var[N]},\nonumber
\end{equation}
where $\langle N^{3}\rangle_{c}$ and $\langle N^{4}\rangle_{c}$ are the third and fourth order cumulants of the multiplicity distribution~\cite{Asakawa:2015ybt}.
The related quantities skewness $S$ and kurtosis $\kappa$ are defined as:
\begin{equation}
	S=\frac{\langle N^{3}\rangle_{c}}{(Var[N])^{3/2}}=\frac{\langle N^{3}\rangle_{c}}{\sigma^{3}},
	\quad \kappa=\frac{\langle N^{4} \rangle_{c}}{Var[N]}=\frac{\langle N^{4}\rangle_{c}}{\sigma^{4}},\nonumber
\end{equation}
where $\sigma^{2}$ is the variance of the multiplicity distribution ($Var[N]=\langle N^{2}\rangle_{c}$). Thus
\begin{equation}
S\sigma=\frac{\langle N^{3}\rangle_{c}}{Var[N]}, \quad \kappa\sigma^{2}=\frac{\langle N^{4} \rangle_{c}}{Var[N]}.\nonumber
\end{equation}
\section{Results}
Preliminary results were obtained from p+p data collected in 2009 at 20, 31, 40, 80 and 158 GeV/c beam momenta. Table~\ref{Tab:events} shows the analysis statistics.
	\begin{table}
	\begin{center}
	\begin{tabular}{l|c|c|c|c|c }
	\hline
	$\sqrt{s_{NN}}$ [GeV]	&	6.3	&	7.6	&	8.7	&	12.3	& 17.3\\
	\hline
Events &	0.2M	&	0.9M	&	3.0M	&	1.7M	&	1.6M	\\
\hline
  \end{tabular}
  \end{center}
  \caption{Number of p+p events taken in 2009 by the NA61/SHINE experiment.}
  \label{Tab:events}
	\end{table}
The analysis acceptance is the same as used for multiplicity and transverse momentum fluctuation analysis~\cite{Aduszkiewicz:2015jna}. Corrected results refer to inelastic interactions and particles produced in strong and electromagnetic processes within the analysis acceptance.

As in Ref.~\cite{Aduszkiewicz:2015jna} multiplicity distributions were corrected for	
	\begin{itemize}
			\item off-target interactions
			\item detector effects
			\item event selection (trigger bias and analysis procedure)
			\item track selection within the analysis acceptance
			\item contribution of weak decays
			\item secondary interactions

	\end{itemize}
Statistical uncertainties were calculated using the sub-sample method\footnote{Statistical uncertainties are smaller than the marker size in the figures}. Systematic uncertainties were estimated by varying event and track selection criteria. 
\begin{figure}[tb]
\centerline{%
\includegraphics[width=.45\textwidth]{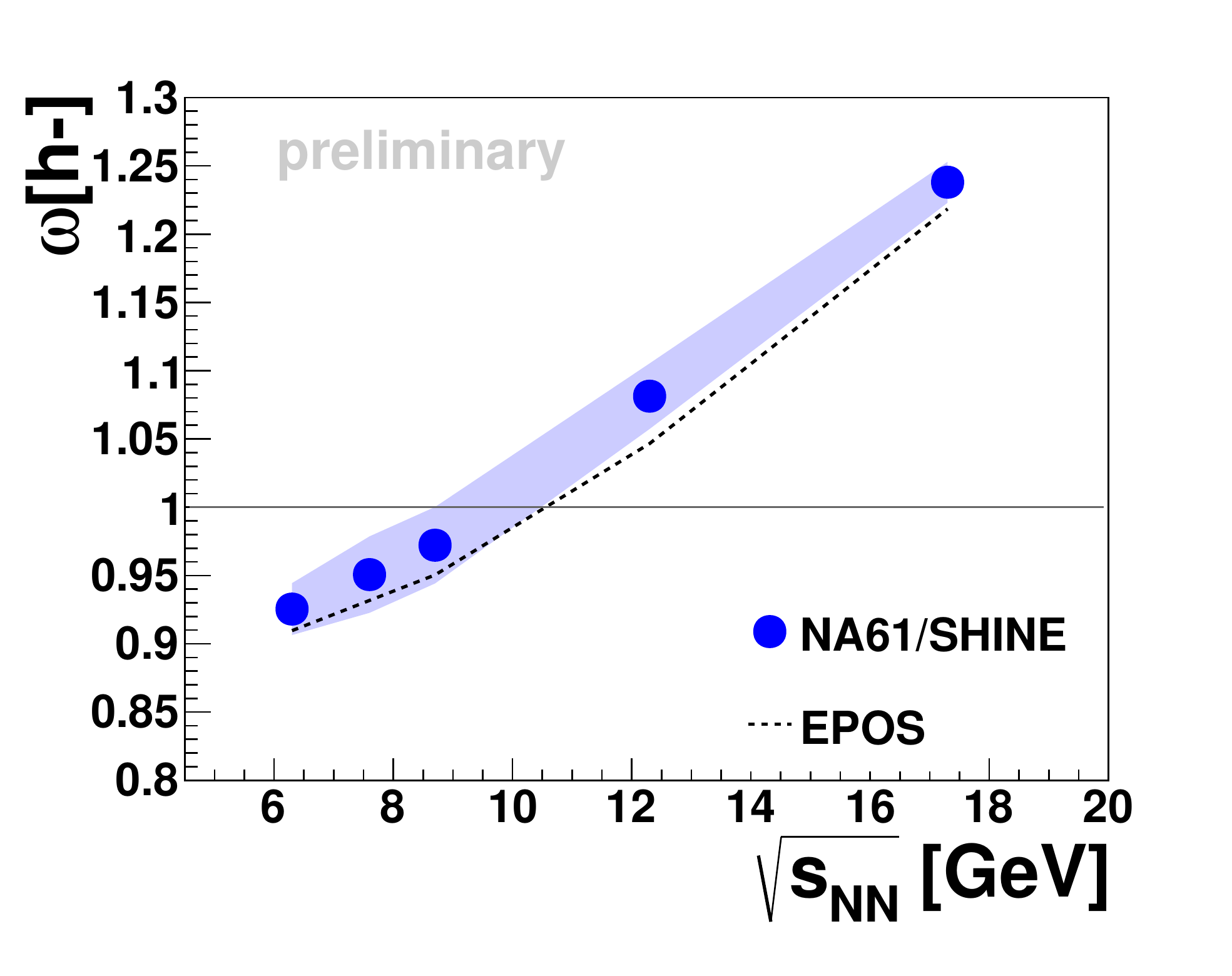}
\includegraphics[width=.45\textwidth]{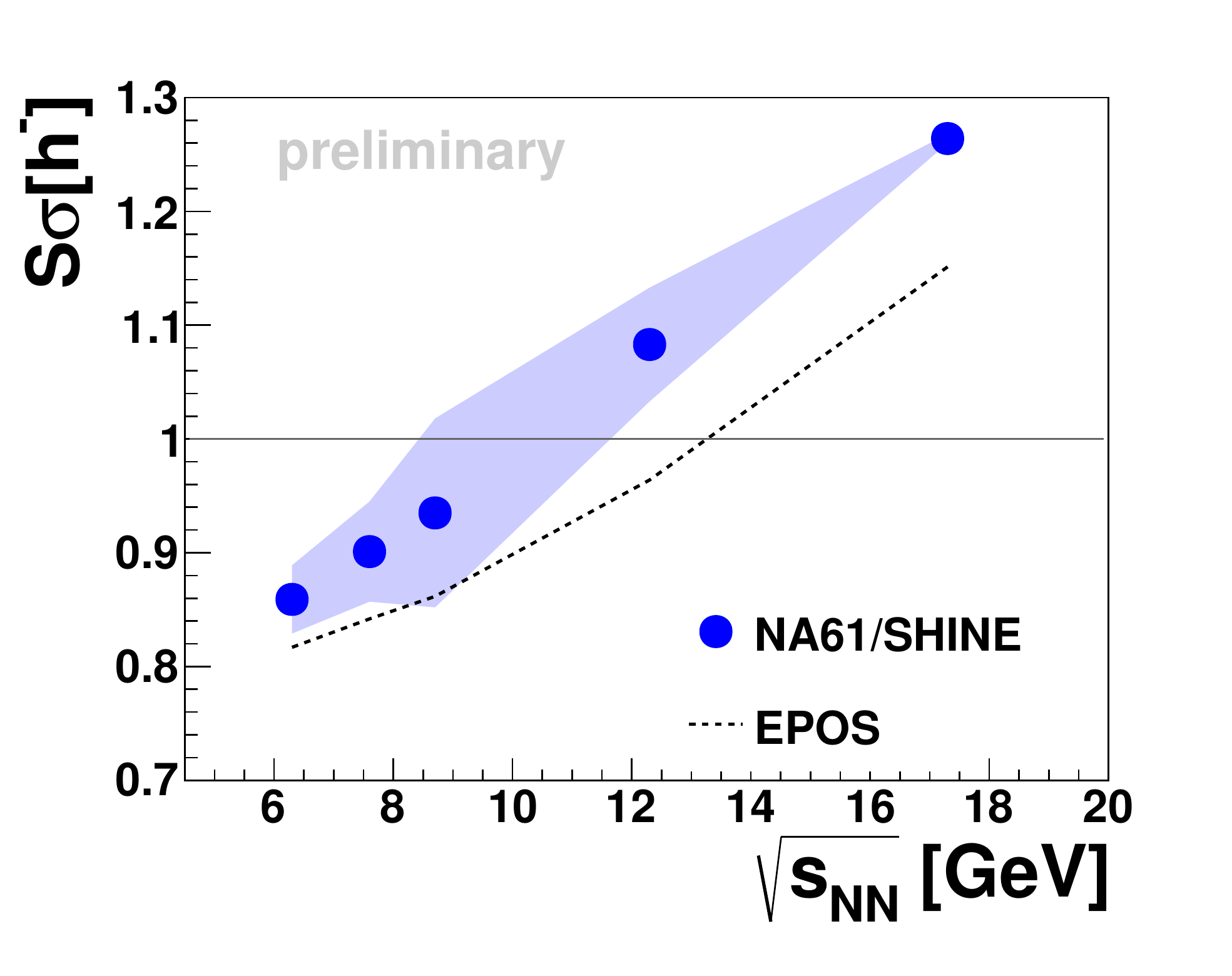}\newline
		\includegraphics[width=.45\textwidth]{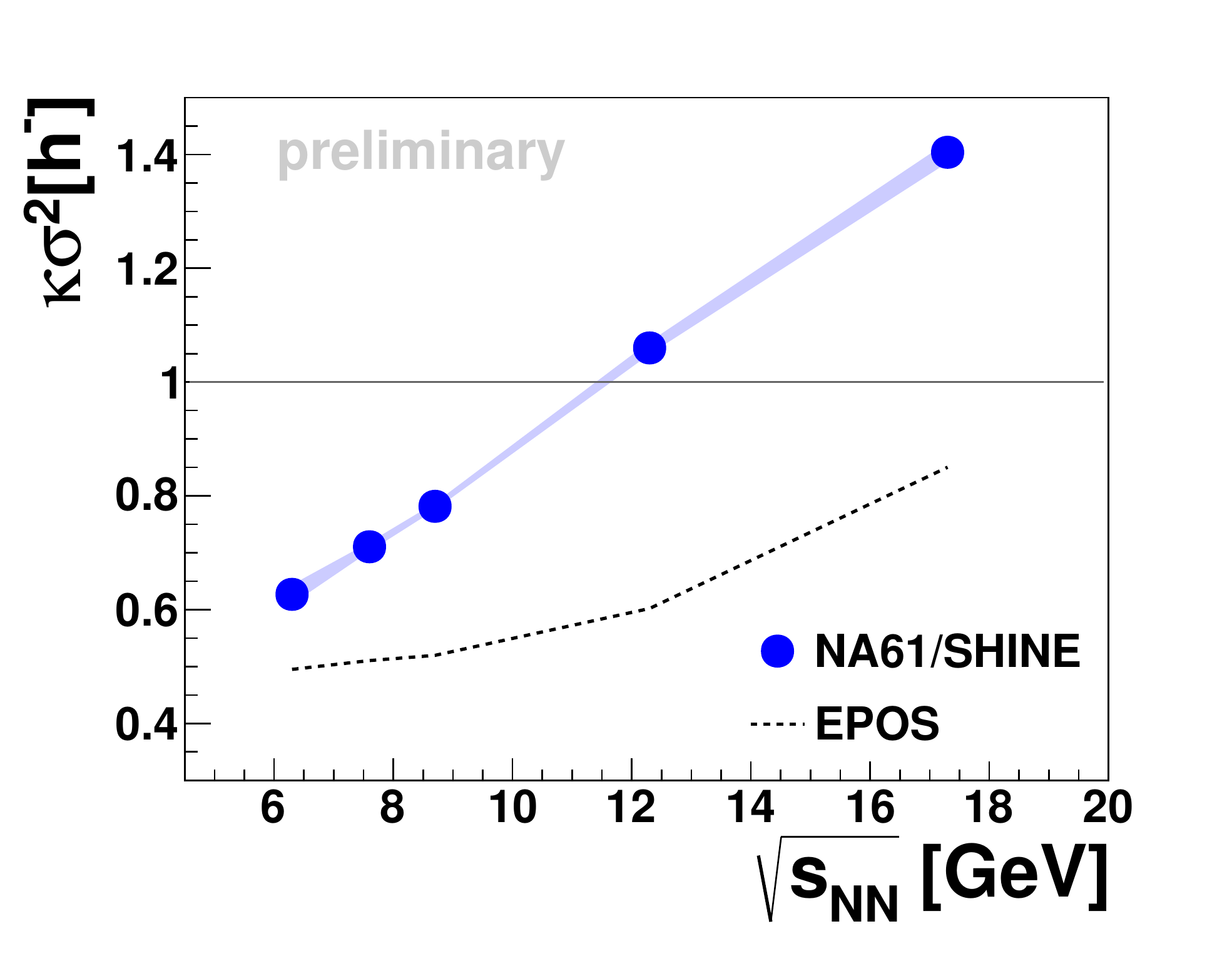}
}
\caption{The energy dependence of $\omega[h^{-}]$, $S\sigma[h^{-}]$ and $\kappa\sigma^{2}[h^{-}]$ in p+p interactions.}
\label{Fig:neg}
\end{figure}
\begin{figure}[tb]
\centerline{%
\includegraphics[width=.45\textwidth]{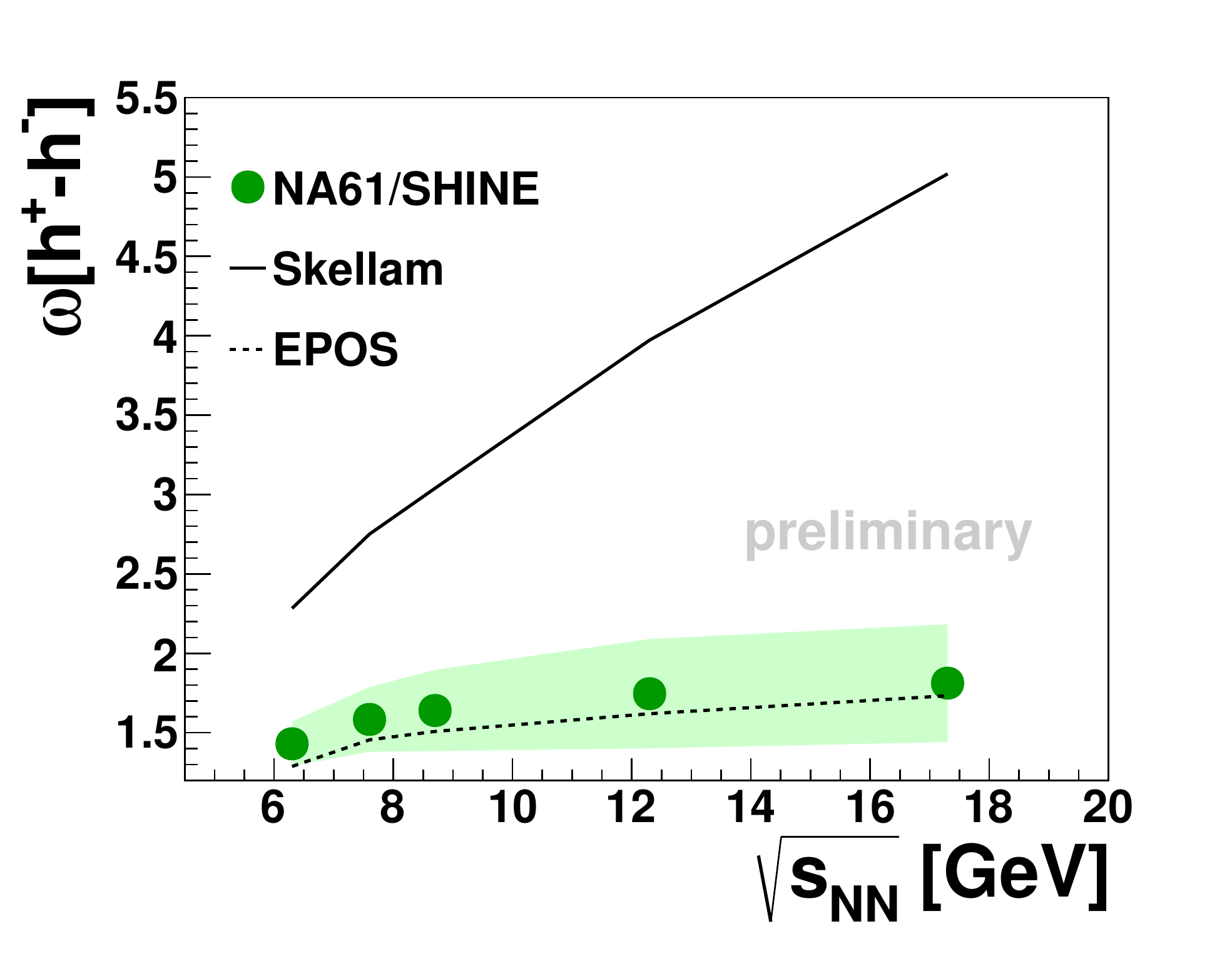}
\includegraphics[width=.45\textwidth]{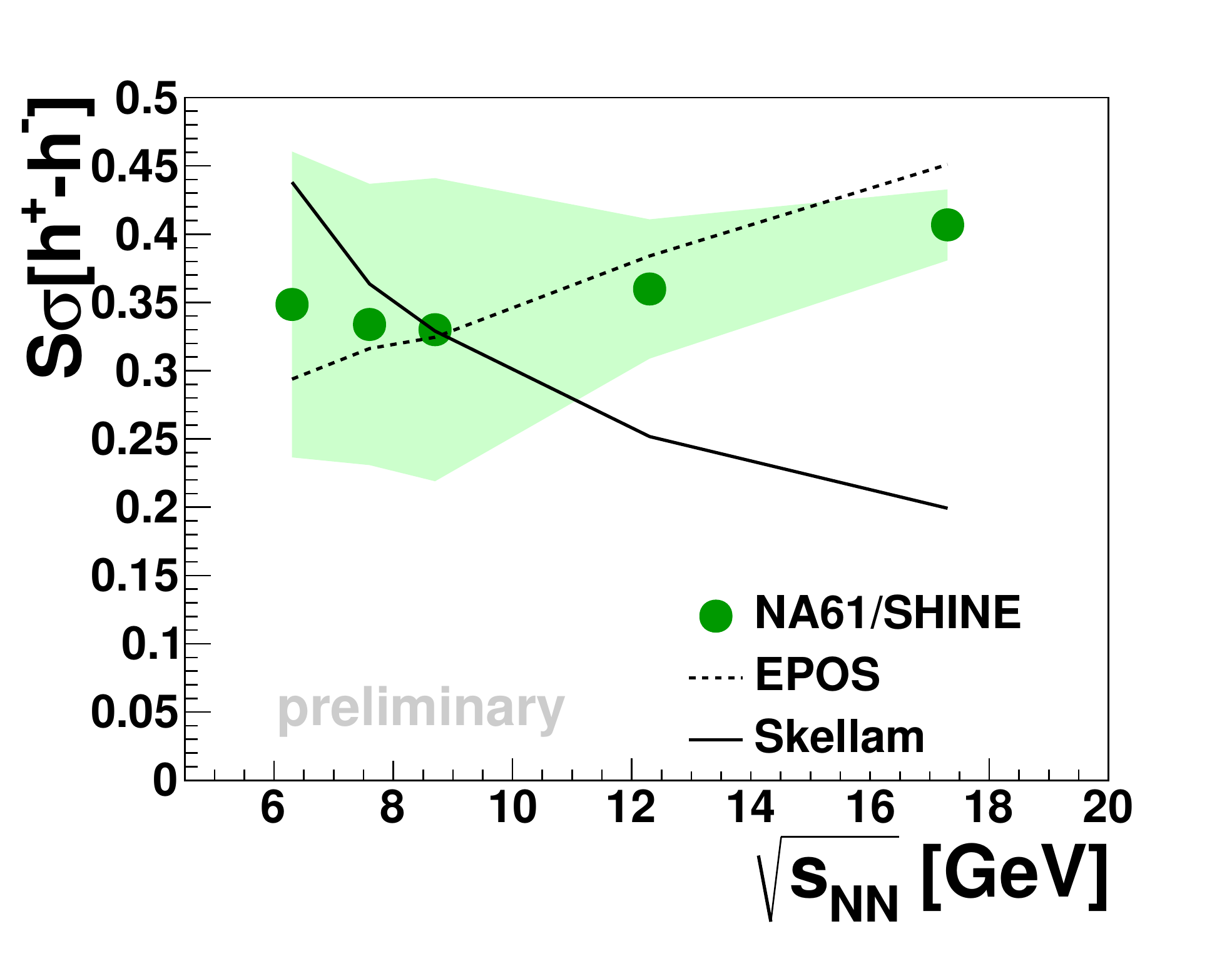}\newline
		\includegraphics[width=.45\textwidth]{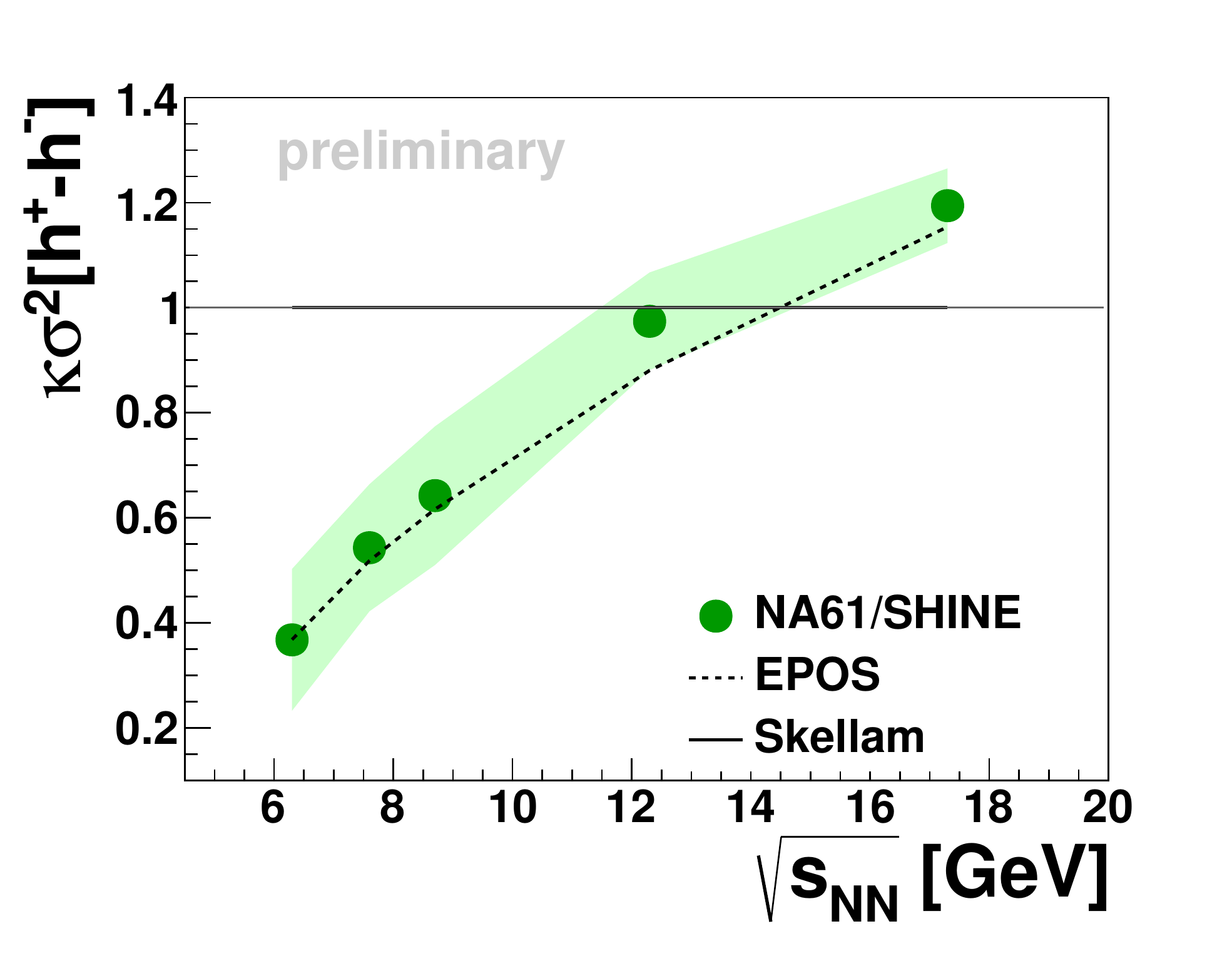}
}
\caption{The energy dependence of $\omega[h^{+}-h^{-}]$, $S\sigma[h^{+}-h^{-}]$ and $\kappa\sigma^{2}[h^{+}-h^{-}]$ in p+p interactions.}
\label{Fig:net}
\end{figure}

Figure~\ref{Fig:neg} shows results on fluctuations of negatively charged hadron multiplicity. The measures $\omega[h^{-}]$, $S\sigma[h^{-}]$ and $\kappa\sigma^{2}[h^{-}]$  rise with collision energy and cross 1 between 40 and 80 GeV/c. These results are not reproduced by statistical models (GCE or CE)~\cite{BegunCPOD2016} (for details see conference slides). 

Figure~\ref{Fig:net} shows results on fluctuations of net-charge. The scaled variance, $\omega[h^{+}-h^{-}]$, as well as $S\sigma[h^{+}-h^{-}]$ depends very weakly on collision energy whereas $\kappa\sigma^{2}[h^{+}-h^{-}]$  rises with collision energy and crosses 1 at 80 GeV/c. 
Net-charge fluctuations measured by $\omega[h^{+}-h^{-}]$ are smaller than predictions of the independent particle production model for which the  net-charge distribution is described by the Skellam distribution. The energy dependences of  $S\sigma$ and $\kappa\sigma^{2}$ are completely different from those predicted by the model. 

The EPOS 1.99 model describes the observed values of $\omega[h^{-}]$, $S\sigma[h^{-}]$ and net-charge fluctuations but it underestimates the value of $\kappa\sigma^{2}[h^{-}]$.

{\small {\bf Acknowledgments:}
This  work  was  partially  supported  by  the  National  Science
Centre, Poland grant 2015/18/M/ST2/00125.}

\end{document}